\documentclass{iaus}
\usepackage{graphicx}

\title[Infrared observations of UCHII regions] 
{Near and Mid Infrared Observations of Ultracompact HII Regions}

\author[Crowther, P.A.]   
{Paul A. Crowther$^1$}

\affiliation{$^1$Department of Physics \& Astronomy, 
University of Sheffield\break Hounsfield Road, Sheffield S3 7RH, UK
\break email: Paul.Crowther@sheffield.ac.uk}

\pubyear{2005}
\volume{227}  
\pagerange{}
\date{14-Jun-05}
\jname{Proceedings IAU Symposium 227, Massive Star Birth: A Crossroads of Astrophysics}
\editors{R. Cesaroni, E. Churchwell, M. Felli \& C.M. Walmsley, eds.}
\begin{document}

\maketitle

\begin{abstract}
This review discusses  near- and mid- infrared observations
of Ultracompact (UC) HII regions. The importance of 
{\it ISO} mid-IR fine-structure nebular lines
 is emphasised, since only a small fraction of UCHII regions are 
observed directly in the near-IR.  The reliability of contemporary 
atmospheric models for such indirect diagnostics  is
discussed, whilst a revised spectral type-temperature calibration is 
presented for Galactic O3 to B3 dwarfs. In particular, fine-structure 
line derived properties of G29.96--0.02 differ from the direct near-IR
spectroscopic result and represents a serious discrepancy which needs to
be addressed. Mid-IR {\it MSX} and {\it Spitzer} imaging permits the  
identification of those UCHII  regions  for which  far-IR {\it IRAS} fluxes 
are 
reliable, relevant to  the single versus cluster nature of individual 
sources. High spatial resolution imaging
with ground-based 8m telescopes allows more direct tests, as  recently 
applied to G70.29+1.60.  Finally, recent {\it Spitzer} mid-IR  
observations of Giant HII regions are briefly discussed.
\keywords{stars: 
formation, ISM: HII regions, infrared: stars}
\end{abstract}

\section{Introduction}

The formation and early evolution of massive stars remains one of the
outstanding puzzles in astrophysics. Despite the ubiquitous presence
of massive stars in galaxies, conventional theory fails to explain how
these stars are born.
Ultracompact HII (UCHII) regions represent the  first evolutionary
stage of massive stars in which their Lyman continuum radiation escapes in
the form of free-free emission at cm wavelengths. Consequently, UCHII regions
connect the early `hot core' stage to later phases when the ionizing source
reveals itself optically. UCHII regions possess high
densities of typically 10$^{4}$ cm$^{-3}$  and
have yet to break out of their natal dust cocoons. Consequently,
the ultraviolet radiation from their hot, massive stars is re-radiated in 
the  thermal infrared. It is widely acknowledged that massive stars do not 
form in isolation. However, individual star  forming regions may host a 
number of UCHII regions, which may be dominated by one massive star rather
than a cluster. This review presents recent observations of UCHII regions
at near- and mid- infrared wavelengths, which aim to address this question.

\begin{table}
\begin{center}
\caption{Spectral type--temperature calibration for Galactic O and early B 
dwarfs, updated from the recent compilation of ionizing fluxes
in the H\,{\sc i} (Q$_0$) and He\,{\sc i} (Q$_1$) ionizing continua by Martins et al. (2005,
their Table~4) to allow for the theoretical mass--radius 
relationship, and extend the sequence to later subtypes: B0--2 results from 
Kilian et al.  (1994) and  Grisby et al. (1992) plus B2--3 results from Morossi 
\&  Malagnini (1985). B star ionizing fluxes are from WM-basic models 
(Pauldrach et al. 2001;  Smith et al. 2002).  Typically, O star 
temperatures are revised downward 
by $\sim$10\%  relative to Vacca  et al. (1996) such that an 
N(LyC)=10$^{49}$ at O6V rather than O7V. Note that recent results for Magellanic 
Cloud OB stars suggest higher stellar temperatures than Milky Way 
counterparts (Heap et al. 2005; Massey et al. 2005)} 
\label{table1} \begin{tabular}{lllrccccc}\hline
Dwarf & $T_{\rm eff}$ & $\log g$ & $R$ & $\log L/L_{\odot}$&M&$M_{\rm 
V}$&log $Q_0$&log $Q_1$ \\
Sp Type &  kK           & cgs & $R_{\odot}$ &   &  $M_{\odot}$    & 
mag         & ph/s      & ph/s \\ \hline
O3      & 45     & 4.0 & 13.8 & 5.85  & 69      & --5.8 & 49.7 & 49.1\\%
O4      & 43     &     & 12.3 & 5.67  & 57      & --5.5 & 49.4 & 48.8\\
O5      & 41     &     & 11.1 & 5.50  & 48      & --5.2 & 49.2 & 48.5\\
O6      & 39     & 4.05& 10.0 & 5.32  & 41      & --4.9 & 49.0 & 48.3\\
O7      & 37     &     & 9.4  & 5.17  & 37      & --4.7 & 48.8 &47.9\\
O8      & 35     &     & 8.9  & 5.03  & 34      & --4.5 & 48.5 &47.4\\
O9      & 33     &     & 8.4  & 4.88  & 32      & --4.3 & 48.1 & 46.3\\
O9.5    & 31.5   & 4.1 & 8.3  & 4.78  & 30      & --4.1& 47.9  & 45.8\\
B0      & 29.5   &     & 7.8  & 4.65  & 28     & --4.0 &  47.4  & 45.2\\
B0.5    & 28     &     & 7.4 & 4.48 & 26       & --3.7 & 46.8  & 44.3\\
B1      & 26     &     & 6.6 & 4.26  & 22      & --3.3 & 46.3  & 43.9\\
B1.5    & 24     & 4.15 & 5.7 & 3.98   & 17      & --2.8& 46.0 & 43.6 \\
B2      & 21     &      & 5.3 & 3.69 & 15      & --2.4& 45.3 & 42.5\\ 
B2.5    & 19     & 4.2  & 4.8 & 3.43 & 13      & --2.0 &44.7 & 41.5 \\
B3    & 17.5     &      & 4.1 & 3.15 & 10      & --1.5& 44.1 & 40.4 \\
\hline
\end{tabular}
\end{center}
\end{table}

\section{Near-IR observation of UCHII regions}


Hanson et al. (2002) carried out a H-band imaging survey of 63 radio-selected
UCHII regions, of which only a small fraction have near-IR 
counterparts, with typical extinctions of $A_{\rm V} \sim$30--50 mag.
Of course, the youngest OB stars are deeply embedded, with perhaps
$A_{\rm V}  \sim$ 1000 mag, so near-IR or even mid-IR 
detection indicates a somewhat more mature ($\sim 10^{5}$ yr) system.
The well known UCHII region G5.89--0.39 represents an excellent case.
The ionizing star is absent in H-band imaging by Hanson et al., whilst
higher spatial resolution K$'$-band imaging by Alvarez et al. (2004) 
reveals a number of  faint sources in the vicinity of the radio peak. 
It is only in the thermal L$'$-band where this source is clearly seen
(Feldt et al. 2003), with L$'$=8.6 mag and K$_s$-L$'$=6.1 mag!


Amongst the small fraction of the ionizing stars that {\it can} be seen in
the K-band, the dominant spectral signature is nebular emission
in Br$\gamma$ and in some cases He\,{\sc i}, with H$_{2}$ usually also seen 
(Hanson et al. 2002). In principal, nebular He\,{\sc i} and Br$\gamma$ emission
lines can be used as indirect probes of the ionizing star (Lumsden et al.
2001), although mid-IR fine structure lines offer a more sensitive
probe, and direct spectroscopic classification is, of course, preferable
to both. Near-IR stellar spectroscopy is more straightforward for sources 
within the same star forming complexes as UCHII regions, albeit away from the
strong nebulosity (Bik et al. 2005).

\begin{figure}[htbp]
\includegraphics[width=14cm]{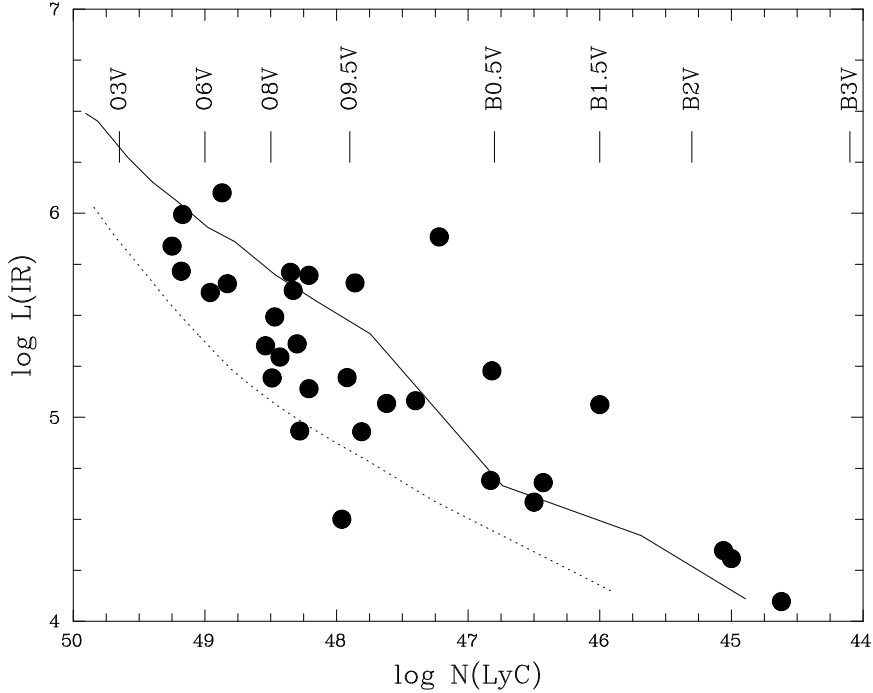}
\caption{Radio derived Lyman continuum fluxes of isolated UCHII
regions, N(LyC), versus bolometric luminosities, L(IR), obtained from mid and far-IR {\it MSX}
and {\it IRAS} fluxes (see Conti \& Crowther 2004), together with predictions
from Lumsden et al. (2003) for a single star (dotted) versus cluster 
(solid, Salpeter IMF) population.}\label{figure}
\end{figure}

To date, the best known 
example of an UCHII region for which the principal ionizing
star can be observed in the near-IR is G29.96--0.02, for which Watson \& 
Hanson (1997) estimated O5--6V from low resolution K-band spectroscopy.
Martin-Hernandez et al. (2003) present high spectral resolution VLT/ISAAC
spectroscopy of G29.96--0.02, confirming a mid O spectral type. Indeed, 
Hanson et al. (these proc.) have presented yet higher quality H and 
K-band  Subaru/IRCS spectroscopy of this source. Quantitative near-IR 
analysis of  this dataset by Hanson et al. reveals $T_{\rm eff} \sim$ 41kK. 
Notably, the wind properties of the ionizing star in G29.96--0.02 appear
to be normal with regard to optically visible OB stars, in contrast with
weak-lined, so-called, OVz stars.

The role of
quantitative near-IR spectroscopy of OB stars has recently been investigated
by Lenorzer et al. (2004) and by Repolust et al. (2005).  The spectral  
type-temperature  calibration of optically visible O stars has been 
significantly revised in recent years  due to the inclusion of line 
blanketing and  stellar winds in atmospheric models (e.g. Crowther
et al. 2002; Massey et al. 2005). A  revision to the recent Martins et 
al. (2005) compilation is indicated in  Table~\ref{table1} to incorporate
early B dwarfs.  Consequently,  $T_{\rm eff}=41$kK for G29.96--0.02 is 
representative of a Galactic O5V, in agreement with  the original 
spectral type of Watson \& Hanson (1997). 

\section{Mid-IR imaging of UCHII regions}

The greatly reduced extinction at mid-IR wavelengths mean that most
UCHII regions can be observed, albeit indirectly. For a 
typical visual extinction  of 50 mag  towards an UCHII region, only 5 mag 
are suffered in the K-band, or 0.25 mag at 25$\mu$m. 

\begin{figure}[htbp]
\includegraphics[width=7cm]{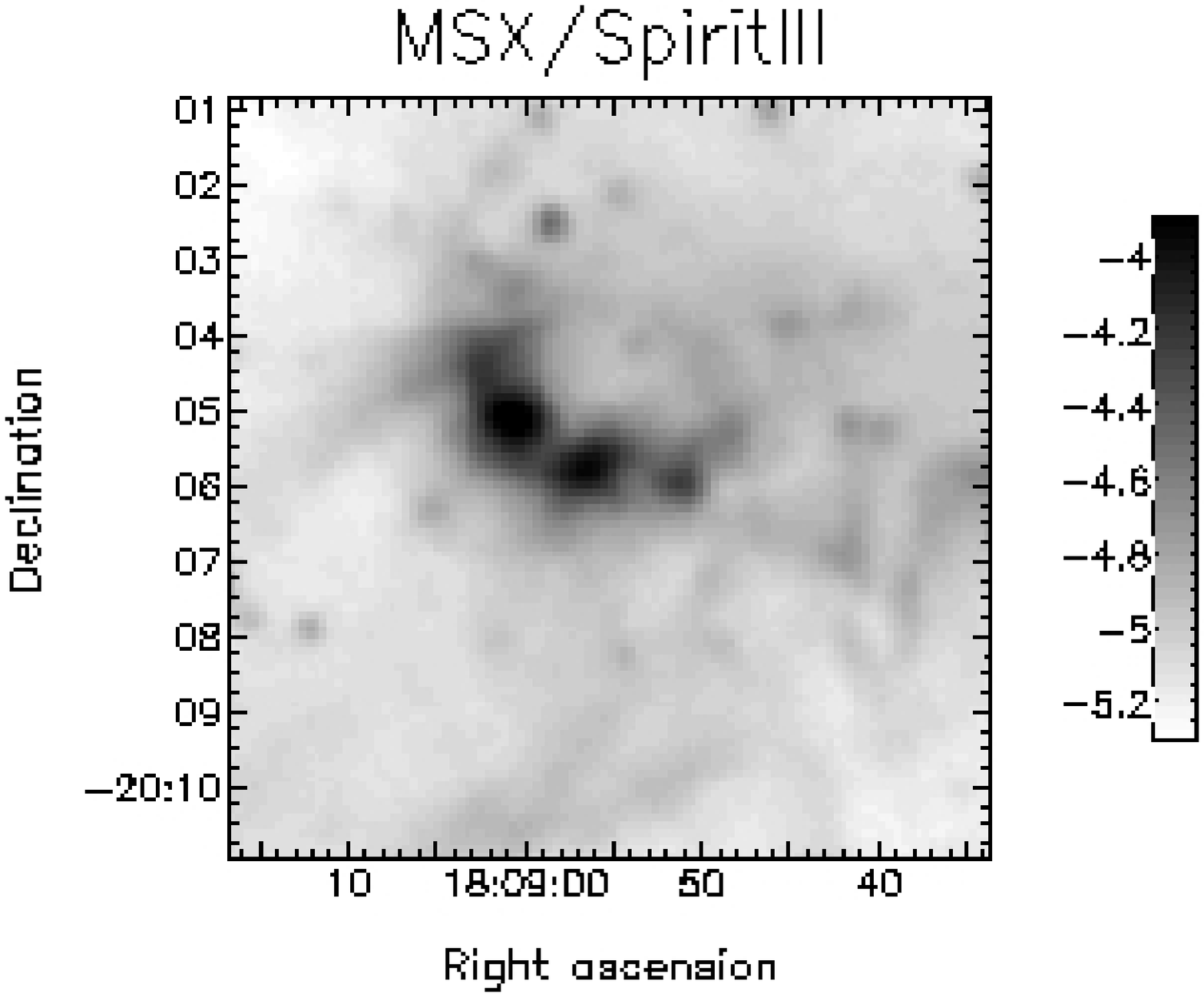}
\includegraphics[width=7cm]{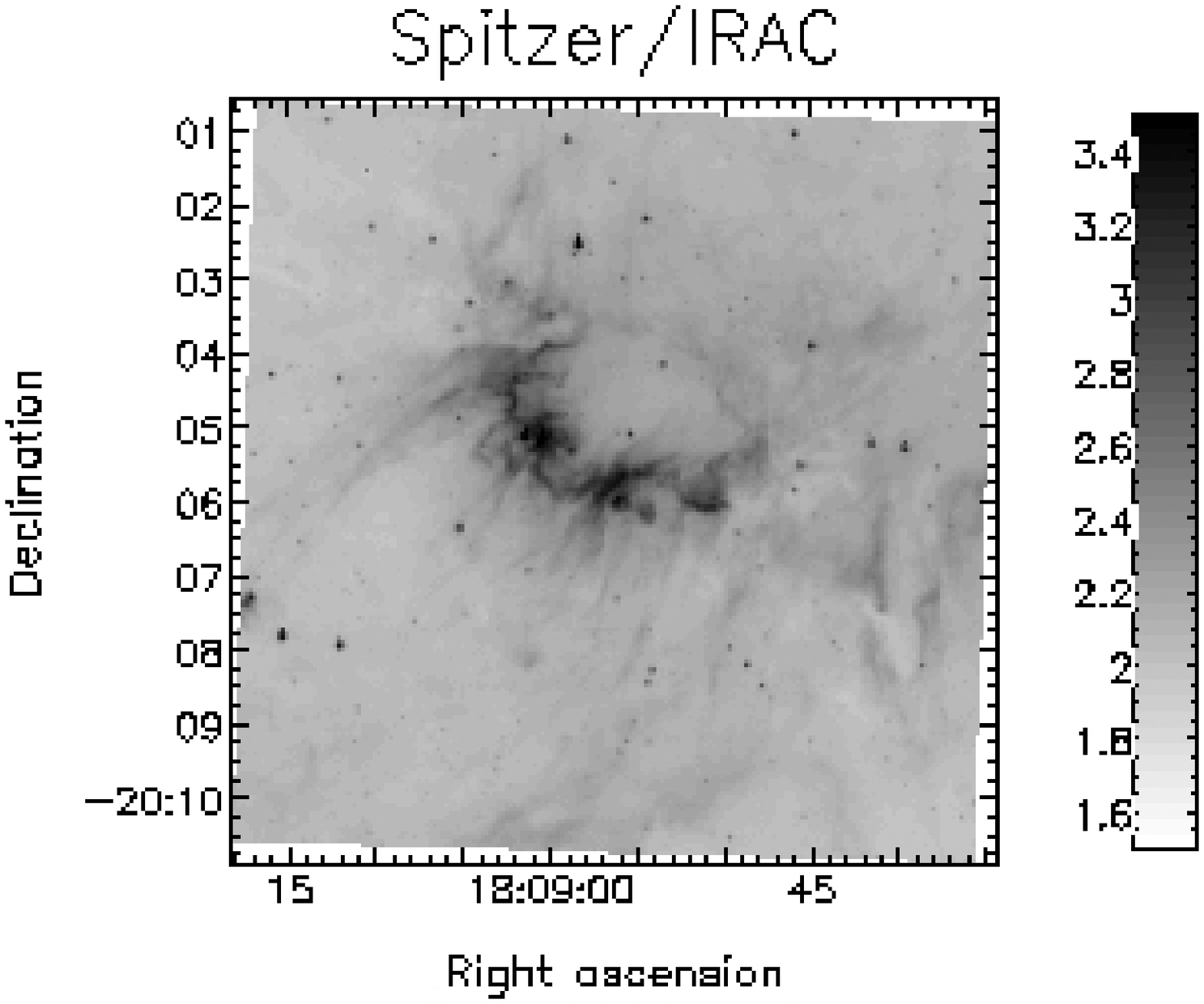}
\caption{Comparison between 8$\mu$m 10x10 arcmin images 
of the UCHII region G10.30--0.15 obtained with {\it MSX} (left)
and the {\it Spitzer} GLIMPSE survey (right). North is up and east 
is to the left in both, logarithmic scale, images}\label{spitzer} 
\end{figure}


The energy emitted from the dust cocoons
surrounding individual UCHII regions peaks in the far-IR, sampled by the 
{\it IRAS} 60 and 100$\mu$m filters. However, the poor spatial resolution of 
{\it IRAS} has severely hindered the interpretation of far-IR fluxes until 
now. Prior to the {\it Spitzer} MIPS far-IR instrument, the Spirit III 
instrument  aboard the {\it Midcourse Space Experiment (MSX)} satellite 
carried out a  mid-IR (8-21$\mu$m) imaging survey of the Galactic plane. 
{\it IRAS} has a 2 arcmin spatial resolution at 100$\mu$m, but has 
nevertheless
been used to obtain an integrated luminosity for individual UCHII regions. 
This has been combined with high spatial resolution (few arcsec) radio
fluxes to investigate the single versus cluster nature of UCHII regions
(e.g. Lumsden et al. 2003).

The large {\it IRAS} far-IR footprint led Crowther \& Conti (2003) to
obtain {\it MSX} imaging of UCHII regions from Wood \& Churchwell (1989) and 
Kurtz et al. (1994) to investigate multiplicity. A mid-IR counterpart was 
identified for 50/51  sources\footnote{One source from the original 
compilations, G5.97-1.17,  is now known as a proplyd (Stecklum et al. 
1998)}, with the exception of G23.26-0.20. The majority of sources were 
detected with {\it MSX} Band C (12$\mu$m), although some were very red (e.g. 
21$\mu$m/12$\mu$m 
$\sim$30 for G45.47--0.05). Unsurprisingly, no correspondence was obtained 
between the moderate resolution (18 arcsec) mid-IR spatial morphologies 
and the 2--6cm VLA datasets at high  resolution (few arcsec). Given et 
al. (2005) have recently provided a much more extensive radio compact 
HII region catalogue, the majority of which are also red in {\it MSX} 
colours.

Better morphological agreement is obtained with respect to lower spatial 
resolution radio observations that are sensitive to the extended 
emission, such as 21\,cm data for G10.30--0.15 by Kim \& Koo (2001). High 
spatial resolution mid-IR datasets closely resemble the radio morphology 
in some cases (e.g.  G29.96--0.02,   Ball et al. 1996), although this is 
not universally true, according to recent ground-based mid-IR imaging (De 
Buizer,  priv. comm.) for which $0.3$ arcsec resolution is achievable.


Approximately 35\% of the UCHII regions were either double or multiple
in the {\it MSX} images, indicating that the {\it IRAS} fluxes are 
significantly 
contaminated by near neighbours in those cases. Although the peak dust 
emission falls longward of the 8--21$\mu$m region, the superior spatial 
resolution of {\it MSX} relative to {\it IRAS} does  allow the  
identification of 
those UCHII regions  for which far-IR fluxes are reliable. In  
Fig.~\ref{figure} we 
compare radio derived Lyman continuum fluxes of isolated UCHII regions 
with bolometric luminosities, together with predictions from Lumsden et 
al. (2003) for a single star  versus a cluster with a Salpeter Initial 
Mass Function (IMF). These predictions used Co-Star non-LTE models 
(Schaerer \& de Koter 1997). Similar predictions are presented by
Sternberg et al. (2003) using WM-basic non-LTE models.

As anticipated, UCHII regions scatter around the cluster case in 
Fig.~\ref{figure}, although
the quoted radio fluxes exclude the extended (15-60 arcsec) emission. 
Observations  which also include this typically lead to a factor of 3 
increase (Kurtz et al. 1999; Kim \& Koo 2001), although recent 
observations by Ellingsen et al. (2005) find more modest increases
in their sample of UCHII regions.  Nevertheless, in all
cases allowance for extended emission shifts individual points to 
towards the single star dominated models.  In addition, dust may absorb 
the ionizing   photons or photons may leak from the region without 
ionizing the surrounding 
material (density rather than ionization bounded case). Consequently,
one has to take care when interpreting these observations.

Of course, {\it Spitzer} now routinely provides a high spatial resolution
view at mid- to far-IR wavelengths with IRAC (4-8$\mu$m) and MIPS (24, 
70$\mu$m). With regard to UCHII regions, the GLIMPSE survey (Benjamin et 
al. 2003) has provided IRAC imaging (e.g. G10.30-0.15 in 
Fig.~\ref{spitzer}), although 24--70$\mu$m 
MIPS images await a Cycle~2 Guest Observer program (P.I. Sean Carey).

\begin{figure}[htbp]
\includegraphics[width=14cm]{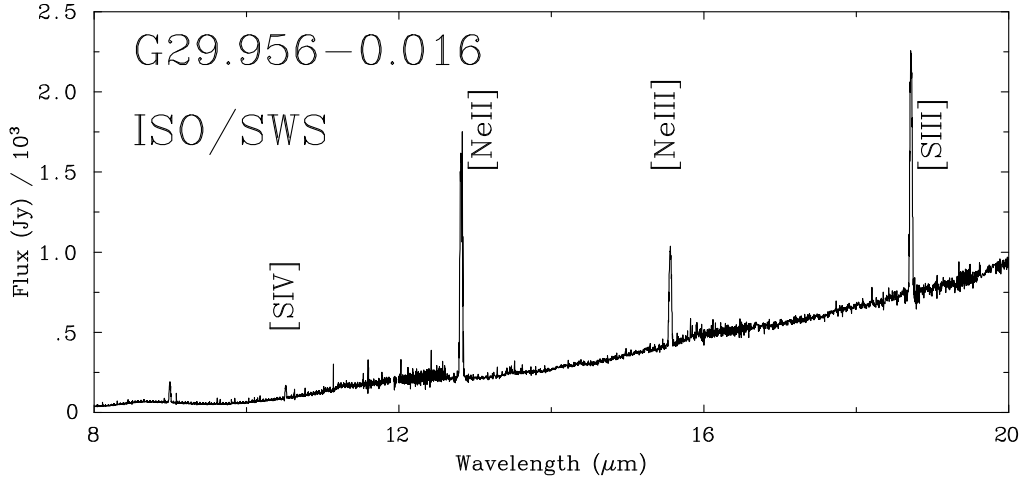}
\caption{Mid-IR ISO/SWS spectrum of G29.96--0.02, revealing 
nebular emission lines superimposed on a hot dust continuum, 
together with dust bands and molecular ices (Peeters et al. 
2002).}\label{g29} \end{figure}

\section{Mid-IR spectroscopy}


The {\it Infrared Space Observatory (ISO)} has made a significant 
contribution to 
the study of UCHII regions spectroscopically in the mid-IR, since important
fine-structure lines (e.g. [NeIII] 15.5$\mu$m, [SIII] 18.7$\mu$m) are 
unavailable from the ground. The {\it ISO} SWS  and LWR instruments were 
well suited to large aperture (14$\times$20 to 
20$\times$33 arcsec) observations of bright sources. Due to spatial crowding,
care has to be taken regarding contamination of {\it ISO} datasets by near 
neighbours. Peeters et al. (2002) present {\it ISO} spectroscopy of UCHII 
regions,
revealing nebular (hydrogen recombination and metallic fine-structure)
Ëemission  lines superimposed upon a hot dust continuum, together with dust
bands (6--13$\mu$m PAH, 9.7$\mu$m silicate) and molecular ices (CO$_2$, 
H$_2$O). A representative spectrum of G29.96--0.02 is presented in 
Fig.~\ref{g29}. Radiative transfer modelling of dust has been discussed by
Faison et al. (1998) and Arthur et al. (2004), whilst 3D modelling is
presented by Whitney (these proc.).

Fine-structure lines originate either from (i) the photo-dissociation region
(PDR), diagnosing the warm gas outside the HII region (e.g. [CII], [OI]), or
(ii) the ionized nebula, providing an indirect tracer of the extreme
ultraviolet (EUV) energy distribution of the OB stars within the 
UCHII region (e.g. [ArII-III], [NeII-III], [SIII-IV]). This
is an extremely useful means of studying the stellar content of the UCHII
region, although different atmospheric models differ in the EUV flux 
distributions. In recent years, significant progress has been made in the 
development of atmospheric models for OB stars. 


Fully-line blanketed LTE  and non-LTE
model atmospheres have been developed by Kurucz (1992) and Lanz \& Hubeny 
(2003), respectively. The latter, TLUSTY, models
are appropriate for stars without stellar winds. Since OB stars do possess
strong stellar winds, either an approximate treatment of metal line 
blanketing is necessary, e.g. CoStar (Schaerer \& de Koter 1997), WM-basic 
(Pauldrach et al. 2001), FASTWIND (Puls et al. 2005), or the metals 
considered are incomplete  e.g. CMFGEN  (Hillier \& Miller 1998).
FASTWIND and CMFGEN currently represent the leading atmospheric
models capable of reproducing the observed UV wind and optical 
photospheric  features simultaneously (e.g. Crowther et al. 2002).

Unfortunately, these non-LTE models disagree with each other at energies
significantly above the Lyman edge (13.6eV), which are sampled by the mid-IR 
fine-structure lines (e.g. Ar$^+$ 28eV, S$^{2+}$ 35eV, Ne$^{+}$ 41eV).
Nevertheless, attempts have been made to derive stellar temperatures from
combining selected non-LTE models with photoionization codes (e.g. 
Giveon et al. 2002; Mokiem  et al. 2003). In particular, from comparison 
between ratios of 
fine-structure lines observed in {\it ISO} spectroscopy of HII regions 
with atmospheric model predictions, Morisset et al. (2004)  concluded
that CoStar models are too hard at high energies due to the approximate
treatment of blanketing, TLUSTY and Kurucz models are too soft at high 
energies due to the neglect of stellar winds, and CMFGEN and WM-basic
models are in reasonable agreement with observations\footnote{FASTWIND 
models are not widely available, although Puls et al. (2005) conclude
their EUV energy distributions are similar to WM-basic.}, though fare no
better than blackbody distributions across selected energy intervals! 
This is of course, far from satisfactory such that further 
improvements in model atmospheres are necessary.


One may attempt to estimate the stellar temperature of
the most massive star in UCHII regions from photoionization modelling
of mid-IR fine structure lines. Amongst the most detailed studies carried out
to date was that of Morisset et al. (2002) who derived T$_{\rm 
eff} = 35\pm$3kK for
G29.96--0.02 based on CMFGEN and WM-basic models, together with a two 
component density distribution -- a small, high density region (0.1pc, 
10$^{4.75}$ cm$^{-3}$) surrounded by an extended halo (1pc, 10$^{2.8}$ 
cm$^{-3}$).

The nebular derived stellar temperature corresponds to an O8$\pm$1.5 dwarf
(Table~\ref{table1}). However, the ionizing star of G29.96--0.02 has
been observed in the K-band revealing an O5V star with $T_{\rm eff}=$41kK
(Hanson et al, these proc.). This represents a serious discrepancy, 
and needs to be resolved in order to progress our understanding of UCHII 
regions from mid-IR spectroscopy. Ideally, the EUV energy distribution 
derived for the ionizing 
star should be combined with a photoionization model to reproduce the 
near and mid-IR nebular emission. Clumping in HII regions doubtless
also plays a role (Ignace \& Churchwell 2004). Whatever the cause,
we cannot hope to understand  more complex cases, such as the integrated 
mid-IR spectra from starburst
galaxies (Verma et al. 2003; Schaerer et al., these proc.), until 
we have resolved with such apparently simple cases as G29.96--0.02.


Recently, Okamoto et al. (2003) have presented high resolution 
Subaru observations of G70.29+1.60 (K3-50A), revealing the first direct 
evidence for a cluster, since fine-structure maps of [ArIII], [SIV] and 
[NeII] indicates multiple sources, one mid- and two late- type O stars. 
The presence of multiple sources in G70.29+1.60 was supported by 
K$'$-band speckle imaging by Hofmann et al. (2004). Clearly, this
technique has great promise for the near future.

\section{Mid-IR observations of Giant HII (GHII) regions}

For OB stars formed in a cluster, their combined Lyman continuum 
luminosity can ionize a substantial volume, resulting in a GHII region
($> 10^{50}$ Lyman photons/s). Near-IR imaging and spectroscopy of GHII regions 
has been discussed by Damineli et al. (these proc.),
and individual {\it Spitzer} IRAC images have been presented by 
Churchwell et al. (2004) for RCW49, Chini (these proc.) for M17 and  
Brandl (these  proc.) for NGC3603 and W49. Conti \&  Crowther (2004) have 
obtained a revised catalogue of $\sim$50 Galactic GHII regions, and compared 
their {\it MSX} derived mid-IR properties with those of UCHII regions. In 
contrast with UCHII regions, where dust intercepts most stellar 
radiation, GHII regions are no longer optically thick so some radiation 
escapes.

{\it Spitzer} spectroscopy and imaging of one GHII region, G10.2--0.3 within 
the 
W31 complex, has recently been obtained in a Cycle 1 Guest Observer program
(P.I. Crowther). Amongst GHII regions, W31 is notable since it contains
(i) massive young stellar objects without radio emission, (ii) UCHII regions 
and (iii)
'naked' O stars (Blum et al. 2001).  {\it Spitzer} IRS spectroscopy reveals 
the 
characteristic mid-IR fine structure lines in individual sources. By way of 
example, [NeIII]  15.5$\mu$m/[NeII]  12.8$\mu$m = 0.3 for one of the 
naked O stars, W31-3, suggesting a mid-O spectral type. Qualitatively, this 
agrees with  that obtained directly from near-IR spectroscopy of W31-3 by 
Blum et al. (2001). Although a detailed study remains to be carried out,
W31-3 indicates a consistent K-band stellar versus mid-IR nebular result,
in contrast with G29.96--0.02, so it is unclear at present which is the
more typical.

Finally, whilst UCHII regions represent the nearby Galactic counterparts
of so-called extragalactic ultra-dense HII (UDHII) regions (Johnson, these 
proc.),  Westerlund~1 (Clark et al. 2005) is currently the best candidate 
for a Galactic  counterpart to extragalactic Super Star Clusters (SSCs).
These are much  more massive and denser than normal Galactic GHII regions 
and are  common in the starburst  galaxies M82 and the Antennae, 
but were hitherto unknown in the Milky Way. In contrast with UCHII and 
GHII regions,  Westerlund~1 is sufficiently mature  to have already 
shed  its natal dust cocoon  and ionized gas, such that it is faint at radio 
wavelengths, and is  only bright  in the near and mid-IR due to its 
stellar content which  includes red supergiants and yellow hypergiants. 
Other examples undoubtedly await detection in the Milky Way.


\begin{acknowledgments}
Thanks to Peter Conti for introducing the author to the wonderful world
of massive star formation in Boulder, Martin Cohen for bringing {\it MSX} to 
my attention, and the Royal Society for continuing financial support.
\end{acknowledgments}


\begin{thebibliography}{}

\bibitem[]{} Alvarez C., Feldt M., Henning T., Puga E., Brandner W., 
2004, ApJS 155, 123
\bibitem[]{} Arthur S.J., Kurtz S.E., Franco J., Albarran M.Y., 2004, ApJ 
608, 282
\bibitem[]{} Ball R., Meixner M.M., Keto E., Arens J.F., Jernigan J.G., 
1996, AJ 112, 1645
\bibitem[]{} Benjamin R.A., Churchwell E., Babler B.L. et al. 2003,  PASP 
115, 943
\bibitem[]{} Bik A., Kaper L., Hanson M.M., Smits, M., 2005, A\&A in 
press (astro-ph/0505293)
 \bibitem[]{} Blum R.D., Damineli A., Conti P.S., 2001, AJ 121, 3149
\bibitem[]{} Churchwell E., Whitney B.A., Babler B.L. et al. 2004, ApJS 
154, 322
\bibitem[]{} Clark J.S., Negueruela I., Crowther P.A., Goodwin S.P., 
2005, A\&A 434, 949
 \bibitem[]{}Conti P.S, Crowther P.A., 2004, MNRAS 355, 899
\bibitem[]{}Crowther P.A., Conti P.S., 2003, MNRAS 343, 143
\bibitem[]{} Crowther P.A., Hillier D.J., Evans C.J. et al. 2002, ApJ 
579, 774
\bibitem[]{} Ellingsen S.P., Shabala S.S., Kurtz S.E., 2005, MNRAS 357, 1003
\bibitem[]{} Faison M., Churchwell E., Hofner P., Hackwell J., Lynch 
D.K., Russell R.W., 1998, ApJ 500, 280
\bibitem[]{} Feldt M., Puga E.,, Lenzen R. et al. 2003, ApJ 599, L91
\bibitem[]{} Giveon U., Sternberg A., Lutz D., Feuchtgruber H., Pauldrach 
A.W.A., 2002, ApJ 566, 880
\bibitem[]{} Giveon U., Becker R.H., Helfand D.J., White R.L., 2005, AJ 
129, 348
\bibitem[]{} Grigsby J.A., Morrison N.D., Anderson L.S., 1992, ApJS 78, 205
\bibitem[]{} Hanson M.M., Luhman K.L., Rieke G.H., 2002, ApJS 138, 35
\bibitem[]{} Heap S.R., Lanz T., Hubeny I., 2005, ApJ in press (astro-ph/0412345) 
\bibitem[]{} Hillier D.J., Miller D.L., 1998 ApJ 496, 407
\bibitem[]{} Hofmann K.-H., Balega Y.Y., Preibisch T., Weigelt G., 2004, 
A\&A 417, 981
\bibitem[]{} Ignace R., Churchwell E., 2004, ApJ 610, 351
\bibitem[]{} Kilian J., Montenbruck O., Nissen P.E., 1994, A\&A 284, 437
\bibitem[]{} Kim K.-T., Koo B.-C.,  2001, ApJ 549, 979
\bibitem[]{} Kurucz R., 1991, in Precision Photometry: Astrophysics
of the Galaxy, Philip A.G.D., Upgren A.R., Janes K.A. (eds), L. Davis Press
\bibitem[]{} Kurtz S.E., Churchwell E., Wood D.O.S., 1994, ApJS 91, 659
\bibitem[]{} Kurtz S.E., Watson A.M., Hofner P., Otte B., 1999, ApJ 514, 232
\bibitem[]{} Lanz T., Hubeny I., 2003, ApJS 146, 417
\bibitem[]{} Lenorzer A., Mokiem M.R., de Koter A., Puls J., 2004, A\&A 
422, 275
\bibitem[]{} Lumsden S.L., Puxley P.J., Hoare M.G., 2001, MNRAS 320, 83
\bibitem[]{} Lumsden S.L., Puxley P.J., Hoare M.G., Moore T.J.T., Ridge 
N.A., 2003, MNRAS 340, 799
\bibitem[]{} Martin-Hernandez M.L., Bik A., Kaper L., Tielens A.G.G.M.,
Hanson M.M., 2003, A\&A 405, 175
\bibitem[]{} Martins F., Schaerer D., Hillier D.J., 2005, A\&A 436, 1049
\bibitem[]{} Massey P., Puls J., Pauldrach A.W.A., Bresolin F., 
Kudritzki R.P., Simon T. 2005, ApJ in press (astro-ph/0503464)
\bibitem[]{} Mokiem M.R., Martin-Hernandez M.L., Lenorzer A., de Koter 
A., Tielens A.G.G.M., 2004, A\&A 419, 319
\bibitem[]{} Morossi C., Malagnini M.L., 1985, A\&AS 60, 365
\bibitem[]{} Morisset C., Schaerer D., Martin-Hernandez M.L. et al.
2002, A\&A 386, 558
\bibitem[]{} Morisset C., Schaerer D., Bouret J.-C., Martins F., 2004, 
A\&A 415, 577 
\bibitem[]{} Okamoto Y.K., Kataza H., Yamashita T. et al. 2003, ApJ 584, 368
\bibitem[]{} Pauldrach A.W.A., Hofmann T.L., Lennon M., 2001, A\&A 375, 161
\bibitem[]{} Peeters E., Martin-Hernandez M.L., Damour F, et al. 2002, 
A\&A 381, 571
\bibitem[]{} Puls J., Urbaneja M.A., Venero R. et al. 2005, A\&A 435, 669
\bibitem[]{} Repolust T., Puls J., Hanson M.M., Kudritzki R.-P., Mokiem 
M.R., 2005 A\&A in press (astro-ph/0501606)
\bibitem[]{} Schaerer D., de Koter A., 1997, A\&A 322, 598
\bibitem[]{} Smith L.J., Noris R.P.F., Crowther P.A., 2002, MNRAS 337, 1309
\bibitem[]{} Stecklum B., Henning T., Feldt M. et al. 1998, AJ 115, 767
\bibitem[]{} Sternberg A., Hoffmann T.L., Pauldrach A.W.A., 2003, ApJ 
599, 1333
\bibitem[]{} Vacca W.D., Garmany C.D., Shull J.M., 1996, ApJ 460, 914
\bibitem[]{} Verma A., Lutz D., Sturm E. et al. 2003, A\&A 403, 829
\bibitem[]{} Watson A.M., Hanson M.M., 1997, ApJ 490, L165
\bibitem[]{} Wood D.O.S., Churchwell E.D., 1989, ApJS 69, 831
\end{thebibliography}
\end{document}